\documentclass[12pt]{iopart}
\usepackage{amssymb,amsfonts}

\def\theequation{\arabic{section}.\arabic{equation}}

\begin{document}

\title{Brane Cosmology with Matter in the Bulk (I)}
\author{Pantelis S. Apostolopoulos$^{\dag}$ and Nikolaos Tetradis$^{\dag}$}

\begin{abstract}
We derive exact solutions of the Einstein equations 
in the context of the Randall-Sundrum model with matter both on the
brane and in the bulk. The bulk metric is a generalization of the static
metrics describing the interior of stellar objects. 
We study the cosmological evolution on the brane. 
Our solutions describe energy outflow from the brane, and the appearance of
``mirage'' contributions to the Hubble expansion that alter 
the standard cosmological evolution.
\end{abstract}

\address{\dag University of Athens, Department of Physics, Nuclear and
Particle Physics Section, Panepistimiopolis, Zografos 15771, Athens, Greece.}
\ead{papost@phys.uoa.gr; ntetrad@cc.uoa.gr}

\maketitle 

\relax
\renewcommand{\theequation}{\arabic{section}.\arabic{equation}}
\def\be{\begin{equation}}
\def\ee{\end{equation}}
\def\bs{\begin{subequations}}
\def\es{\end{subequations}}

\def\lx{\lambda}
\def\Lx{\Lambda}

\newcommand{\tl}{\tilde t}
\newcommand{\ttt}{\tilde T}
\newcommand{\rhot}{\tilde \rho}
\newcommand{\ptt}{\tilde p}
\newcommand{\drho}{\delta \rho}
\newcommand{\drhot}{\delta {\tilde \rho}}
\newcommand{\dchi}{\delta \chi}

\newcommand{\rhb}{\tilde \rho}
\newcommand{\pb}{\tilde p}

\newcommand{\A}{A}
\newcommand{\B}{B}
\newcommand{\mmu}{\mu}
\newcommand{\mnu}{\nu}
\newcommand{\ii}{i}
\newcommand{\jj}{j}
\newcommand{\jr}{]}
\newcommand{\ml}{\sharp}

\newcommand{\da}{\dot{a}}
\newcommand{\db}{\dot{b}}
\newcommand{\dn}{\dot{n}}
\newcommand{\dda}{\ddot{a}}
\newcommand{\ddb}{\ddot{b}}
\newcommand{\ddn}{\ddot{n}}
\newcommand{\pa}{a^{\prime}}
\newcommand{\pn}{n^{\prime}}
\newcommand{\ppa}{a^{\prime \prime}}
\newcommand{\ppb}{b^{\prime \prime}}
\newcommand{\ppn}{n^{\prime \prime}}
\newcommand{\fda}{\frac{\da}{a}}
\newcommand{\fdb}{\frac{\db}{b}}
\newcommand{\fdn}{\frac{\dn}{n}}
\newcommand{\fdda}{\frac{\dda}{a}}
\newcommand{\fddb}{\frac{\ddb}{b}}
\newcommand{\fddn}{\frac{\ddn}{n}}
\newcommand{\fpa}{\frac{\pa}{a}}
\newcommand{\fpb}{\frac{\pb}{b}}
\newcommand{\fpn}{\frac{\pn}{n}}
\newcommand{\fppa}{\frac{\ppa}{a}}
\newcommand{\fppb}{\frac{\ppb}{b}}
\newcommand{\fppn}{\frac{\ppn}{n}}

\newcommand{\rht}{\tilde{\rho}}
\newcommand{\wt}{\tilde{w}}

\newcommand{\dA}{\dot{A_0}}
\newcommand{\dB}{\dot{B_0}}
\newcommand{\fdA}{\frac{\dA}{A_0}}
\newcommand{\fdB}{\frac{\dB}{B_0}}

\def\be{\begin{equation}}
\def\ee{\end{equation}}
\def\bs{\begin{subequations}}
\def\es{\end{subequations}}
\newcommand{\een}{\end{subequations}}
\newcommand{\ben}{\begin{subequations}}
\newcommand{\beq}{\begin{eqalignno}}
\newcommand{\eeq}{\end{eqalignno}}
\def \lta {\mathrel{\vcenter
     {\hbox{$<$}\nointerlineskip\hbox{$\sim$}}}}
\def \gta {\mathrel{\vcenter
     {\hbox{$>$}\nointerlineskip\hbox{$\sim$}}}}

\def\g{\gamma}
\def\mpl{M_{\rm Pl}}
\def\ms{M_{\rm s}}
\def\ls{l_{\rm s}}
\def\l{\lambda}
\def\m{\mu}
\def\n{\nu}
\def\a{\alpha}
\def\b{\beta}
\def\gs{g_{\rm s}}
\def\d{\partial}
\def\co{{\cal O}}
\def\sp{\;\;\;,\;\;\;}
\def\r{\rho}
\def\dr{\dot r}
\def\dt{\dot\varphi}
\def\e{\epsilon}
\newcommand{\NPB}[3]{\emph{ Nucl.~Phys.} \textbf{B#1} (#2) #3}   
\newcommand{\PLB}[3]{\emph{ Phys.~Lett.} \textbf{B#1} (#2) #3}   
\newcommand{\ttbs}{\char'134}           
\newcommand\fverb{\setbox\pippobox=\hbox\bgroup\verb}
\newcommand\fverbdo{\egroup\medskip\noindent%
                        \fbox{\unhbox\pippobox}\ }
\newcommand\fverbit{\egroup\item[\fbox{\unhbox\pippobox}]}
\newbox\pippobox
\def\tr{\tilde\rho}
\def\lb{w}
\def\bbox{\nabla^2}
\def\mt{{\tilde m}}
\def\rct{{\tilde r}_c}
\def \lta {\mathrel{\vcenter
     {\hbox{$<$}\nointerlineskip\hbox{$\sim$}}}}
\def \gta {\mathrel{\vcenter
     {\hbox{$>$}\nointerlineskip\hbox{$\sim$}}}}

\section{Introduction}
\setcounter{equation}{0}

The idea that our Universe is a defect embedded in a higher-dimensional
bulk space \cite{rubakov} provides a context in which the 
cosmological evolution can display novel features. 
Of particular interest is the possibility of exchange of energy between
the defect and the bulk space. The first obstacle in realizing this idea
is the requirement to localize low-energy gravity on the defect. This is
necessary in order to reproduce the correct form of Newton's law, and 
conventional cosmological evolution with the Hubble parameter depending 
linearly on the energy density of the matter localized on the defect.
The Randall-Sundrum model \cite{rs} provides an example in which
this requirement is satisfied. The defect is a four-dimensional hypersurface
(a 3-brane) in a five-dimensional bulk with negative cosmological
constant (AdS space). The geometry is non-trivial (warped)
along the fourth spatial dimension, so that an effective localization of
low-energy gravity takes place near the brane. The matter is assumed to be
concentrated only on the brane.
For low matter
densities the cosmological evolution on the brane
becomes typical of a Friedmann-Robertson-Walker Universe \cite{binetruy,csaki}.

It is possible to discuss the brane evolution either in a coordinate system
(system A) in which the brane is located at a fixed value of the fourth spatial
coordinate and the bulk is time-dependent \cite{binetruy,csaki}, or
in a different coordinate system (system B) 
in which the bulk is static and the brane
is moving \cite{kraus}. In the latter case, the bulk metric is 
five-dimensional AdS-Schwarzschild \cite{birm}. The two points of view are
equivalent \cite{gregory}. Several aspects of the cosmological evolution 
have been the subject of a large number of studies. (For recent reviews,
with extensive lists of references, see ref. \cite{brax}.)

We are interested in the possibility of energy exchange between the brane
and the bulk, and the modifications it would induce on the cosmological
evolution on the brane. For such a study 
the bulk must be assumed to contain, apart from the cosmological constant,
some matter component.
The basic equations have been studied in the 
coordinate system A. However, in this system
it is difficult to obtain an exact solution for the bulk metric. 
In the presence of bulk matter, the cosmological evolution on the brane
is not autonomous, and the explicit knowledge of the bulk energy-momentum
tensor is necessary. As a result, certain assumptions often
need to be made for its form \cite{vandebruck}--\cite{tet2}. 
Previous studies have considered a bulk field \cite{dilaton}, or
radiation emitted by the brane into the bulk \cite{hebecker,vaidya}.
Also an attempt has been made to incorporate thermal effects in the bulk
\cite{rothstein}.

In this study we start by deriving an explicit solution for the bulk in
the coordinate system B. We allow for a negative cosmological constant and
additional matter with an arbitrary equation of state. 
In the following section we obtain a particular class of static solutions, 
that correspond to generalizations of the AdS-Schwarzschild metric.
These solutions can be interpreted as AdS-stars, as their form
resembles the solution for the interior of stellar configurations.
In section 3 we introduce a brane as the boundary of the 
five-dimensional space. We employ the coordinate system A, in which 
the brane is located at a fixed value of the fourth spatial coordinate.
We also establish the connection between the two coordinate systems.
This permits us to derive the exact equations governing the cosmological
evolution on the brane, making use of the explicit solutions of section 2.
In section 4 we study the cosmological evolution on the brane. The solutions
describe energy outflow from the brane and the presence of
``mirage'' contributions to the Hubble expansion.
In the final section we summarize our results.

\section{The bulk}
\setcounter{equation}{0}

We consider an action of the form
\be
S=\int d^5x~ \sqrt{-g} \left( M^3 R +\Lambda +{\cal L}_B^{mat}\right)
+\int d^4 x\sqrt{-\gamma} \,\left( -V+{\cal L}_b^{mat} \right), 
\label{001}
\ee
where $R$ is the curvature scalar of the five-dimensional metric 
$g_{AB}, A,B=0,1,2,3,4$,
$-\Lambda$ the bulk cosmological constant ($\Lx >0$),
${\gamma}_{ab}$, $a,b=0,1,2,3$,
the induced metric on the 3-brane, and  
$V$ the brane tension.

We assume that the metric in the bulk can be put in the static form
\begin{equation}
ds^{2}=-n^{2}(r) dt^{2}+r^{2}d\Omega_k^2
+b^{2}(r)dr^{2},
\label{metric}
\end{equation}
where $d\Omega^2_k$ is the metric in a
maximally symmetric three-dimensional space. 
We use $k=-1,0,1$ to parametrize the spatial curvature.
The non-zero components of the five-dimensional Einstein tensor are
\begin{eqnarray}
{G}^{\,0}_{~0} &=&  \frac{3}{b^2} \frac{1}{r}
\left( \frac{1}{r}-\frac{b'}{b} \right) - \frac{3k}{r^2}
\label{ein00} \\
 {G}^{\,i}_{~j} &=& 
\frac{1}{b^2}\left[
\frac{1}{r}\left( \frac{1}{r}+2\frac{n'}{n}\right) 
-\frac{b'}{b}\left( \frac{n'}{n}+2\frac{1}{r}\right)
+\frac{n''}{n} \right]
-\frac{k}{r^2}
\label{einij} \\
{G}^{\, 4}_{~4} 
&=& \frac{3}{b^2}\frac{1}{r}\left(\frac{1}{r}+\frac{n'}{n} \right)
-\frac{3k}{r^2},
\label{ein44} 
\end{eqnarray} 
where primes indicate derivatives with respect to
$r$.

The Einstein equations take the usual form 
\begin{equation}
G^{A}_{~B}
= \frac{1}{2 M^3} T^{A}_{~B} \;, 
\label{einstein}
\end{equation}
where $T^A_{~B}$ denotes the total energy-momentum tensor.
Assuming a negative cosmological constant $-\Lx$ and a 
static perfect fluid in the bulk, 
we write the bulk energy-momentum tensor as
\begin{equation}
T^A_{~B}=
{\rm diag}(\Lambda-\rho,\Lambda+p,\Lambda+p,\Lambda+p,\Lambda+p).
\label{tmn4}  
\end{equation}
Combining the 00 and 44 Einstein equations we obtain
\be
\frac{(bn)'}{bn}=\frac{1}{6M^3}\,b^2 r\, (\rho+p).
\label{extra1} \ee
>From the conservation of the energy-momentum tensor 
we have
\be
\frac{n'}{n}=-\frac{p'}{\rho+p}.
\label{extra2} \ee

In order to solve the Einstein equations we proceed in complete analogy to
the standard study of stellar structure \cite{weinberg}. We rewrite the 
00 Einstein equation as
\be 
\left(\frac{r^2}{b^2} \right)'=2kr+\frac{1}{3M^3}r^3(\Lx-\rho).
\label{ijre} \ee
This has the solution
\be
\frac{1}{b^2}=k+\frac{1}{12M^3}\Lx r^2 - \frac{1}{6\pi^2 M^3}\frac{1}{r^2}
{\cal M}(r),
\label{sol1} \ee
where ${\cal M}(r)$ satisfies
\be
\frac{d{\cal M}}{dr}=2\pi^2 r^3 \rho.
\label{sol1b} \ee
For $\rho=0$ and constant ${\cal M}$ 
we obtain the known AdS-Schwarzschild solution
\cite{birm}.
Through the use of 
eqs. (\ref{extra2}), (\ref{sol1}), the 44 Einstein equation
becomes
\be
\frac{p'}{\rho+p}=\frac{1}{r}-\frac{1}{r}
\left[k+\frac{1}{6M^3}r^2(p+\Lx) \right]
\left[k+\frac{1}{12M^3}\Lx r^2 
- \frac{1}{6\pi^2 M^3}\frac{1}{r^2}{\cal M}(r)\right]^{-1}.
\label{sol2} \ee
The solution of eqs. (\ref{extra2}), (\ref{sol1})--(\ref{sol2}) automatically
satisfies eqs. (\ref{einij}), (\ref{extra1}) that were not used in the
derivation.

For a complete solution the equation of state 
$p=p(\rho)$ must be specified. Then,
if initial conditions are given for $\rho$ and ${\cal M}$ at some $r$,
the system of equations
(\ref{sol1b}), (\ref{sol2}) can be integrated.
The resulting solution 
determines $b(r)$ through substitution 
in eq. (\ref{sol1}). Finally, the integration of eq. (\ref{extra1}) or  
(\ref{extra2}) determines $n(r)$.

The existence of a solution depends crucially on our assumption for the form 
of the metric. 
In the absence of matter in the bulk, 
the most general solution is the AdS-Schwarzschild that
can be put in the form of the ansatz 
(\ref{metric}). This solutions results from
our equations under the assumption $\rho=p=0$: 
Eq. (\ref{extra1}) gives $b=1/n$, while
the integration of eq. (\ref{ijre}) gives eq. (\ref{sol1}) 
with constant $\cal{M}$. 

In the presence of matter in the bulk,
the ansatz of eq. (\ref{metric}) is not the most general.
However, as we shall show in the following, our assumption 
leads to a physically motivated solution, that provides
an interesting example of cosmological evolution induced on the brane. 
Physically, the distribution of matter in the bulk corresponds to that 
of a stellar object in AdS space. Most of the matter is concentrated
near the center of the star at $r=0$, but a non-zero density exists for large
$r$ as well. In this sense, the brane moves within a background that can
be characterized as an AdS-star configuration. 
If all the matter falls into a black-hole
singularity at $r=0$, our solution reduces to the 
standard AdS-Schwarzschild metric in five dimensions.

In a pure AdS background, described by eq. (\ref{metric}) 
with $n^2(r)=1/b^2(r)=k+\Lx r^2/(12 M^3)$, 
the concentration of matter is expected to 
be enhanced near $r=0$. We are interested in
solutions with finite energy density throughout the bulk.
For this reason we consider the boundary 
condition $\rho(r=0)=\rho_0$, $p(r=0)=p_0$. The solution of
eq. (\ref{sol2}) is now uniquely specified. 
For the solution of eq. (\ref{sol1b}) we assume that ${\cal M}(r=0)=0$.
In this way ${\cal M}(r)$ can be roughly interpreted as the integrated matter
between $r=0$ and some arbitrary value of $r$.
In order to prevent the last term in brackets in
the r.h.s. of eq. (\ref{sol2}) from vanishing at a certain $r$, we are
forced to take $k=1$. This implies
that the brane Universe must have positive spatial curvature. The functions
$1/b^2(r)$ and $n^2(r)$ never vanish. 
The resulting solution is finite everywhere in the bulk and has no horizons.
For the cases $k=0,-1$ the function $1/b^2(r)$ vanishes at a certain $r$, which
forces $n^2(r)$ to vanish at the same point, as can be seen from eq. (\ref{extra1}). 
As a result, our ansatz assumes a static
distribution of matter in the presence of horizons in
the metric. This requirement cannot be satisfied, as the divergence of $p$ implied
by eq. (\ref{sol2}) indicates.
It is probable that an alternative non-static
ansatz for the metric (such as the generalization of
the Vaidya metric \cite{vaidya}) must be considered.

The solution of eqs. (\ref{sol1b}), (\ref{sol2}) with $k=1$ has strong similarities 
with the well known solutions for the interior of stellar configurations.
We cannot provide an analytical solution for all $r$, but we check our conclusions
through the numerical integration of eqs. (\ref{sol1b}), (\ref{sol2}).
For a generic polytropic equation of state $p=w \rho^\gamma$
and the boundary conditions we chose,
the solution is regular for small $r$ with $\rho=\rho_0$, 
${\cal M}(r)\sim \rho_0 r^4$.\footnote{ 
For $\gamma=1$  there is another solution that diverges
near $r=0$ according to $\rho\sim r^{-2}$, ${\cal M}\sim r^{2}$. 
This is similar to the standard neutron star solution \cite{weinberg}.} 
The large-$r$ behaviour can be obtained analytically through the assumption
that the cosmological constant term $\sim \Lx r^2$
dominates over the curvature and matter terms. In this way
we obtain
\begin{eqnarray}
&\rho(r)=\left(\kappa \, r^{\frac{1-\gamma}{\gamma}} 
-\frac{1}{w}\right)^{\frac{1}{\gamma-1}} 
~~~~~~~~~~~~~~~~~~~
&{\rm for}~~\gamma\not=1
\label{sol11} \\
&\rho(r)=\kappa\,r^{-\frac{w+1}{w}}~~~~~~~~~~~~~~~~~~~~~~~~~~~~~~&{\rm 
for}~~\gamma=1,
\label{sol12} \end{eqnarray}
with $\kappa$ an integration constant.
Regular behaviour near $r=0$ is always connected with this type of asymptotic behaviour,
as we have verified numerically.

Let us consider first the large-$r$ behaviour for 
$\gamma\not=1$. For $\gamma<1$ (we always
assume $\gamma>0$),
we obtain $\rho\sim r^{-1/\gamma}$, $p\sim r^{-1}$. The density of bulk matter
is non-zero for all $r$. 
We obtain 
\be
{\cal M}(r)=\frac{2\gamma\pi^2}{4\gamma-1} \kappa^{\frac{1}{\gamma-1}} 
\, r^{\frac{4\gamma-1}{\gamma}}
+{\cal M}_\kappa,
\label{mr1} \ee
with ${\cal M}_\kappa$ another integration constant. 
For $\gamma>1$ there is a finite value of $r_s$ for
which $\rho=p=0$. At this point eq. (\ref{extra1}) implies that
$b(r_s)=1/n(r_s)$. We can assume that the matter density remains zero for
$r>r_s$. This permits the matching of the solution for $r<r_s$ with the
standard AdS-Schwarzschild metric for $r>r_s$. Simply, in the latter
we must take ${\cal M}={\cal M}(r_s)$. This is analogous to
the matching of the interior solution with the Schwarzschild metric at the
surface of a star.

In cosmological applications the linear equation of state
with $\gamma=1$ is most commonly employed. 
For large $r$ and $w>0$ the matter density never becomes exactly zero, even though it 
falls off as a power of $r$. 
We obtain 
\be
{\cal M}(r)=\frac{2w\pi^2}{3w-1} \kappa \, r^{\frac{3w-1}{w}}
+{\cal M}_\kappa,
\label{mr2} \ee
with ${\cal M}_\kappa$ an integration constant. 
For the particular value $w=1/3$ the asymptotic behaviour is
\be
{\cal M}(r)=2\pi^2 \kappa \, \ln \left( \frac{r}{r_\kappa}\right)
+{\cal M}_\kappa,
\label{mr3} \ee
with $r_k$ another integration constant.

\section{The brane}
\setcounter{equation}{0}

For the discussion of the cosmological evolution on the brane we follow
ref. \cite{kraus}.
First we consider a coordinate system 
(using Gaussian normal coordinates) in which the metric takes the form
\begin{equation}
ds^2=\gamma_{ab}dx^a dx^b+d\eta^2 
=-\mt^2(\tilde{\tau},\eta )d\tilde{\tau}^2
+{\tilde a}^2(\tilde{\tau},\eta )d\Omega_k ^2+d\eta ^2.
\label{sx2.1}
\end{equation}
The brane is located at $\eta=0$. In order to keep our
discussion simple we identify the half-space $\eta>0$ with the half-space
$\eta<0$, in complete analogy to ref. \cite{rs}.
We define
the proper time on the brane
$d\tau =m(\tilde{\tau},\eta=0)\, d\tilde{\tau}$.
We can rewrite the above metric as
\begin{equation}
ds^2
=-m^2(\tau,\eta )d\tau^2
+a^2(\tau,\eta )d\Omega_k ^2+d\eta ^2,
\label{sx2.1ex}
\end{equation}
with 
$m(\tau,\eta)=\mt(\tilde{\tau},\eta)/\mt(\tilde{\tau},\eta=0)$,
$a(\tau,\eta)={\tilde a}(\tilde \tau,\eta)$,
so that $m(\tau,\eta=0)=1$.

Through an appropriate coordinate transformation 
\begin{equation}
t=t({\tau},\eta ),\qquad r=r({\tau},\eta )  \label{sx2.4}
\end{equation}
the metric (\ref{metric}) 
can be written in the form of eq. (\ref{sx2.1ex}).
It is obvious that $r\equiv a({\tau},\eta )$. Therefore, we obtain
\begin{equation}
dt=\frac{\partial t}{\partial {\tau}}d{\tau}+\frac{\partial t}{
\partial \eta }d\eta  \label{sx2.5}
\end{equation}
\begin{equation}
dr=\frac{\partial r}{\partial {\tau}}d{\tau}+\frac{\partial r}{
\partial \eta }d\eta \equiv \frac{\partial a}{\partial {\tau}}
d{\tau}+\frac{\partial a}{\partial \eta }d\eta .  \label{sx2.6}
\end{equation}
Identification of the metrics (\ref{metric}) and (\ref{sx2.1ex}) requires
\begin{eqnarray}
b^2(r)\left( \frac{\partial a}{\partial {\tau}}\right)^2-n^2(r)\left( 
\frac{\partial t}{\partial {\tau}}\right)^2&=&-m^2({\tau},\eta)
  \label{sx2.8}
\\
b^2(r)\left( \frac{\partial a}{\partial \eta }\right) ^2-n^2(r)\left( \frac{
\partial t}{\partial \eta }\right) ^2&=&1  \label{sx2.9}
\\
b^2(r)\,\,\frac{\partial a}{\partial {\tau}}\frac{\partial a}{\partial
\eta }-n^2(r)\,\,\frac{\partial t}{\partial {\tau}}\frac{\partial t}{\partial
\eta }&=&0.  \label{sx2.10}
\end{eqnarray}

We define $R({\tau})=a({\tau},\eta=0)$. 
In the system of coordinates $(t,r)$ of eq. (\ref{metric})
the brane is moving, as it is located at $r=R({\tau})$. 
At the location of the brane, we find 
\begin{eqnarray}
\frac{\partial t}{\partial \tau} &=& 
\frac{1}{n(R)} \left[ b^2(R) \dot{R}^2 +1 \right]^{1/2}
\label{tr1} \\
\frac{\partial t}{\partial \eta} &=& 
-\frac{b(R)}{n(R)} \dot{R}
\label{tr2} \\
\frac{\partial a}{\partial \tau} &=& 
\dot{R}
\label{tr3} \\
\frac{\partial a}{\partial \eta} &=& 
-\frac{1}{b(R)} \left[ b^2(R) \dot{R}^2 +1 \right]^{1/2},
\label{tr4} \end{eqnarray}
where the dot denotes a derivative with respect to proper time.
The negative sign in the r.h.s. of eq. (\ref{tr4}) is a consequence of
our assumption that $r$ decreases on both sides of the brane
\cite{kraus}.

For completence, we also give the components of the brane velocity 
${\bf u}$ in 
the coordinate system $(t,r)$:
\begin{eqnarray} 
u^t&=&\frac{1}{n(R)}\left[ b^2(R) \dot{R}^2+1 \right]^{1/2}
\label{v1} \\
u^r&=&\dot{R}.
\label{v2} \end{eqnarray}
The unit vector ${\bf n}$ 
normal to the boundary defined by the brane motion has
\begin{eqnarray} 
n^t&=&-\frac{b(R)}{n(R)}\dot{R}
\label{n1} \\
n^r&=&
-\frac{1}{b(R)}\left[ b^2(R) \dot{R}^2+1 \right]^{1/2}.
\label{n2} \end{eqnarray}

It is now straightforward to express the bulk energy-momentum tensor 
at the location of the brane in the coordinate system 
$({\tau},\eta)$. We find
\begin{eqnarray}
T^0_{~0}&=&\Lx-\rho(R)-\Bigl[\rho(R)+p(R)\Bigr]b^2(R)\dot{R}^2
\label{t00} \\
T^1_{~1}=T^2_{~2}=T^3_{~3}
&=& \Lx +p(R)
\label{t11} \\
T^4_{~4}&=&
\Lx+p(R)+\Bigl[\rho(R)+p(R)\Bigr]b^2(R)\dot{R}^2
\label{t44} \\
T^0_{~4}
&=&b(R)\dot{R}\left[ b^2(R) \dot{R}^2+1 \right]^{1/2}
\Bigl[ \rho(R)+p(R)\Bigr].
\label{t04} 
\end{eqnarray}
We have indicated explicitly the dependence of 
$\rho$, $p$ and $b$ on $R$ in order to emphasize that they 
are the solutions of eqs. 
(\ref{sol1})--(\ref{sol2}), with $R$ as their argument.

The equations governing the cosmological evolution on the brane 
can be obtained in various ways. The most straightforward is
to write the Einstein equations in the system $({\tau},\eta)$,
assuming that the brane energy momentum tensor has the form
\be
\ttt^A_{~B}=
\delta(\eta)\,{\rm diag}(V-\rht,V+\tilde{p},V+\tilde{p},V+\tilde{p},0),
\label{branet} \ee
with $V$ the brane tension and $\rht$, $\tilde{p}$ the energy density and pressure
of a perfect fluid localized on the brane.
The details of the calculation can be found in refs. 
\cite{brax}--\cite{tet2} and
will not be repeated here.
The evolution on the brane is governed by the equations
\begin{eqnarray}
\dot{\rht}+3 \frac{\dot{R}}{R}(\rht+\tilde{p})&=&
-2 T^0_{~4}
\label{energ} \\
H^2=\frac{\dot{R}^2}{R^2}&=&\frac{1}{144 M^6}\left(\rht^2+2V\rht\right)
-\frac{k}{R^2}+\chi+\phi+\lx
\label{hubble} \\
\dot{\chi}+4H\chi&=&\frac{1}{36M^6}(\rht+V)T^0_{~4}
\label{chi}\\
\dot{\phi}+4H\phi&=&-\frac{1}{3M^3}H\left( T^4_{~4}-\Lx \right),
\label{phi}
\end{eqnarray}
where $\lx=(V^2/12M^3-\Lx)/12M^3$ is the effective cosmological constant,
and $T^4_{~4}$, $T^0_{~4}$ are given by eqs. (\ref{t44}), (\ref{t04}).
For the particular form of the energy-momentum tensor that we are considering,
we find that 
\be
\chi+\phi=\frac{1}{6 \pi^2M^3}\frac{{\cal M}(R)}{R^4}
\label{chiphi} \ee
satisfies eqs. (\ref{chi}), (\ref{phi}),
if ${\cal M}(R)$ is the solution of eq. (\ref{sol1b}) with $r$ replaced
by $R$.  In order to show this, we make use of  
eq. (\ref{hubble}) in the form
\begin{eqnarray}
\frac{\dot{R}^2}{R^2}=H^2&=&\frac{1}{144 M^6}\left(\rht^2+2V\rht\right)
-\frac{k}{R^2}+\frac{1}{6 \pi^2M^3}\frac{{\cal M}(R)}{R^4}+\lx
\nonumber \\
&=&
\frac{1}{144M^6}\left(\rht+V \right)^2
-\frac{1}{b^2(R)\,R^2}.
\label{hubble2} \end{eqnarray}

We can now express 
eq. (\ref{t04}) as
\be
T^{0}_{~4}
=\frac{1}{12M^3}b^2(R)\,R^2\, H \left( \rht + V \right)
\Bigl[ \rho(R)+p(R)\Bigr].
\label{t04r} 
\ee
The rate of energy trasfer is fixed
by the static solution for the bulk that we derived in section 2. 
This solution is consistent only with energy loss by the brane. 
In order to understand better this point it is convenient to employ the
coordinate system B as in section 2. In this system the brane forms the
moving boundary of a static bulk. As the boundary shifts towards larger values
of the coordinate $r$, energy must be transferred from the brane to the bulk.
Our solution corresponds to the situation that matter leaves the boundary 
fast enough so as to fill the additional volume before 
flows develop from the rest of the bulk space.
Of course there are constraints on the realization of such a scenario.
If the brane matter cannot escape fast enough the bulk cannot remain static.
The detailed discussion of this issue must deal with complicated issues,
such as the exact profile
of the brane beyond the approximation by a $\delta$-function,
the mechanism that localizes particles on the brane, the
dynamics near the surface (pressure gradients, local flows). 
In this work we simply assume that matter
can escape from the brane fast enough for our solution to be realized.

An alternative way to derive eq. (\ref{hubble2}) is by considering
the junction conditions for the boundary defined by the brane motion. 
In the coordinates $({\tau},\eta)$ 
of eq. (\ref{sx2.1ex}) the extrinsic curvature of the 
boundary is
\begin{equation}
K_{ab}=\frac{1}{2}\partial_\eta \gamma_{ab}.
\label{extr} \end{equation}
The junction conditions are \cite{israel}
\be
\Delta K_{ab}=K^+_{ab}-K^-_{ab}=-\frac{1}{2M^3}
\left(T_{ab}-\frac{1}{3}T^c_{~c}\gamma_{ab} \right),
\label{junction} \ee
where $\pm$ denote the two sides of the brane, and the
energy-momentum tensor is given by eq. (\ref{branet}) neglecting the
$\delta$-function.
For the spatial components we find
\be 
K^{\pm}_{ij}=\mp \frac{1}{R}\left(\dot{R}^2
+\frac{1}{b^2(R)}\right)^{1/2}
\gamma_{ij}.
\label{extr1} \ee
Then eq. (\ref{junction}) immediately reproduces eq. (\ref{hubble2}).

\section{The cosmological evolution}
\setcounter{equation}{0}

The cosmological evolution on the brane is described by
eqs. (\ref{energ}), (\ref{hubble2}), with $T^0_{~4}$ given by
eq. (\ref{t04}) or (\ref{t04r}),
and $\rho(R)$, $p(R)$, ${\cal M}(R)$, $b(R)$ determined by the
solution of eqs. (\ref{sol1})--(\ref{sol2}) that we discussed
in section 2. 

The evolution has some characteristic properties: The energy on
the brane is diluted by the expansion, but 
is also reduced through the outflow from both sides
of the brane. This implies that the energy density of the various components 
of brane matter (dust, radiation etc) does not scale as function
of $R$ with the ``naive'' power $R^{-3(1+\wt)}$
determined by the equation of state $\tilde{p}=\wt \rht$.
The outflow makes the reduction faster.

The effective Friedmann equation (\ref{hubble2}) for the expansion 
on the brane has several terms: 
The effective cosmological constant 
$\lx$ leads to exponential expansion. In order to
get conventional cosmology we have to set it to zero
through the appropriate fine-tuning of the bulk cosmogical constant and
the brane tension: $V^2=12M^3 \Lx$.
The curvature term has the standard form. In the context of the
solution of section 2 for the bulk we have $k=1$.
The brane matter induces a term $\sim \rht^2$, characteristic of the
unconventional early cosmology of the Randall-Sundrum model
\cite{binetruy}. At low energy densities $\rht \ll V$ ,
the standard contribution $\sim \rht$ to the Hubble
parameter dominates. The effective Planck constant is
$\mpl^2=12M^6/V$.

The novel feature of eq. (\ref{hubble2}) is the 
term $\sim {\cal M}(R)/R^4$. It is a generalization of the 
``mirage'', or ``Weyl'', or ``dark'' radiation term 
\cite{brax,tet1,hebecker,vaidya,mirage}
encountered in
brane cosmology, which is related to the mass of the black hole
in the bulk. We recover the $1/R^4$ dependence, modulated by the
explicit $R$-dependence of the mass term ${\cal M}$.
Similar modifications have been found in several variations of the
basic Randall-Sundrum scenario \cite{dilaton,vaidya,induced}.

For a bulk described by the solutions of section 2 
the cosmological expansion on the brane at early times 
is dominated by the brane matter density. The reason is that the 
bulk effects are incorporated in the term $\sim {\cal M}(R)/R^4$ that becomes
constant in the limit $R\to 0$. The energy outflow 
$T^0_{~4} = R^2H \rht\left(\rho_0+p_0 \right)/(12M^3)$
also becomes subleading in this limit, and the energy density on the brane
is reduced mainly through the expansion:
$\rht\sim R^{-3(1+\wt)}$.
The scale factor obeys $R\sim t^{1/(3+3\wt)}$, similarly to the 
high-energy regime of the standard Randall-Sundrum model.

We are mainly interested in the large-$R$ behaviour of the cosmological
solution. In this limit, the form of $\rho(R)$ has been determined 
analytically at the end of section 2, and is given by 
eqs. (\ref{sol11}), (\ref{sol12}) with $r$ replaced by $R$.
For $\gamma > 1$ the energy density of bulk matter vanishes for a certain $R$
and remains zero subsequently. The cosmological evolution on the brane
has a complicated behaviour up to a certain time. Later on, it becomes
conventional, with the ``mirage'' radiation term the only indication
of the presence of the bulk.

For $\gamma<1$ and large $R$ we have $\rho(R)\sim R^{-1/\gamma}$, 
$p(R)\sim R^{-1}$.\footnote{We 
emphasize that, for $\gamma\to 1$, the scale factor $R$
must be sufficiently 
large for the term $\sim  R^{(1-\gamma)/\gamma}$ to dominate
over $1/w$ in eq. (\ref{sol11}).} We also obtain $1/b^2=\Lx R^2/(12M^3)$
and $T^0_{~4}=HV(\rho+p )/\Lx$.
In this way we find that for large $R$ the energy density on the 
brane obeys $d(\rht\,R^{3(1+\wt)})/dR\sim -R^{1+3\wt}$. For $\wt \geq 0$ 
this implies
that the energy density becomes zero at some finite $R_f$. Our
solution does not have a physical meaning beyond this point, as it 
describes matter with negative energy density. For $R_f>R\gg 1$
the form of ${\cal M}(R)$, as given by eq. (\ref{mr1}),  
implies that there is a ``mirage'' 
contribution to the Hubble parameter that scales 
$\sim R^{-1/\gamma}$, as well as a ``mirage'' radiation term. 
The former contribution dominates the radiation term
for $\gamma > 1/4$ and sufficiently large $R$.
The influence of the bulk matter
on the brane evolution is equivalent to
that of a ``mirage'' matter component with a linear
equation of state and $\wt_{eff}=-1+1/(3\gamma)$. For $1/3<\gamma<1$, we have
$\wt_{eff}<0$. For $1/2<\gamma<1$ the ``mirage'' term dominates the curvature
term for large $R$. 

The most common form of the equation of state in cosmological applications
has $\gamma=1$. This choice 
leads, through use of eq. (\ref{sol12}),
to $d(\rht\,R^{3(1+\wt)})/dR\sim -R^{1+3\wt-1/w}$.
For $w<1/(2+3\wt)$ the energy density $\rht$ never vanishes, but falls off
asymptotically according to the standard relation $\rht\sim R^{-3(1+\wt)}$.
For $w\geq 1/(2+3\wt)$ 
the energy density $\rht$ becomes zero at a finite $R_f$,
beyond which our solution does not have a physical meaning.
Eqs. (\ref{mr2}), (\ref{hubble2}) imply that 
the influence of the bulk is equivalent to
a ``mirage'' radiation term and another ``mirage'' 
component with 
$\wt_{eff}=-2/3+1/(3w)$. 
For $w>1/2$ we have $\wt_{eff}<0$. For a ``stiff''
equation of state with $w=1$, the ``mirage'' term acts as an additional
curvature term for large $R$.

The most attractive from the above scenaria assumes linear equations
of state for both the bulk and brane matter with $w<1/(2+3\wt)$.
For large $R$ the energy outflow becomes
subleading, and the brane energy density falls off according to
the standard relation $\sim R^{-3(1+\wt)}$. 
For $w<1/3$ the main ``mirage'' effect is a radiation
term. For $w=1/3$ this term gets logaritmic corrections $\sim \ln(R)/R^4$. 
For $1/3<w<1/(2+3\wt)$ the leading ``mirage'' effect 
is a component with an effective equation of state with 
$\wt_{eff}=-2/3+1/(3w)$. In all the scenaria
the ``mirage'' effects are negligible at early times, when the brane energy
density dominates the evolution. However, at later times this energy is
depleted through the expansion and the outflow and the ``mirage'' terms
(as well as the curvature term) dominate the Hubble expansion.

As a final remark we point out that in the limit of vanishing
matter energy density both in the bulk
and the brane our solution reduces to the Randall-Sundrum solution. As a result
we expect localization of the massless Kaluza-Klein mode of the
graviton around the brane. This is
consistent with the emergence of a cosmological evolution typical of a
four-dimensional Universe for brane energy densities smaller than the brane
tension and bulk energy densities smaller than the absolute value of
the cosmological constant.
According to our discussion at the end of section 2, 
the cosmological constant term
always dominates the bulk metric for large values of the scale factor $R$. 
Similarly, the brane tension dominates the brane energy density 
in the same limit.

\section{Conclusions}

The general picture that emerges from our discussion is that 
the presence of matter in the bulk affects the brane evolution in 
unexpected ways. The ``mirage'' radiation term, that has appeared
in previous studies, is only the simplest manifestation of the
existence of the bulk. It comes about because of 
the presence of a black hole at
$r=0$ in an AdS bulk, in the static coordinate system of eq. (\ref{metric}). 
In this study we constructed more general bulk solutions that permit a
non-trivial distribution of bulk matter. Because of the bulk geometry most
of this matter is concentrated in the region near $r=0$ in the 
coordinates of eq. (\ref{metric}), or 
for $\eta \to \infty$ in the coordinates of eq. (\ref{sx2.1ex}).
However, it is possible for the brane to be located in a region of 
non-zero density of bulk matter. As a consequence, the cosmological evolution 
on the brane is modified. 

The first important effect is the flow of
energy away or towards the brane. The solutions we constructed are consistent
only with energy outflow, accompanied with a faster than usual dilution of
brane matter. However, it is possible that more sophisticated bulk solutions 
could permit the influx of energy, accompanied with exciting novel effects
\cite{tet1,tet2}. The second consequence is the modification of the 
standard Hubble expansion through ``mirage'' contributions to the 
brane energy density. The effective equations of state of such components
depend on the equation of state of the bulk matter in a non-linear way.
For example, bulk matter with a linear equation of state and $w>0$ may
lead to a ``mirage'' component on the brane with $\wt_{eff}<0$. 
A general property of the solutions we derived is that the ``mirage'' effects
are negligible at early times when the brane energy density dominates, but
become significant at late times. 

The bulk solutions we derived are only a small sub-class of the
more general solutions that could predict interesting low energy cosmological
evolution for a brane Universe. The novel features 
introduced by the presence of matter in the AdS bulk is the 
outflow or influx of energy from the brane, and the appearance of ``mirage''
components affecting the cosmological evolution on the brane. 
Such features could be useful for modifying the evolution so as to 
include, for example, an accelerating era, as implied by recent astrophysical
observations \cite{perl}. The task of deriving new bulk solutions is difficult,
but the implications appear to be very interesting.

\vskip 0.5cm
\centerline{\bf\large Acknowledgments}
\vskip .3cm

We would like to thank T. Christodoulakis for useful discussions.
The work of N. Tetradis was  partially supported through the RTN contract
HPRN--CT--2000--00148 of the European Union and the research program ``Kapodistrias'' of the Ministry of
National Education and Religious Affairs. Both authors acknowledge financial support through the research program ``Pythagoras'', Grand No 016.

\end{document}